# A comparison of methods for the analysis of binomial proportion data in behavioral research.

Alberto Ferrari, Mario Comelli

*University of Pavia, Department of Behavioral and Brain Sciences, Medical Statistics section*

e-mail: alberto.ferrari04@universitadipavia.it, aferrari34@yahoo.com

tel: +39 0382 98 7534

**Abstract**

In behavioral and psychiatric research, data consisting of a per-subject proportion of "successes" and "failures" over a finite number of trials often arise. This kind of clustered binary data are usually non-normally distributed, which can cause issues with parameter estimation and predictions if the usual general linear model is applied and sample size is small.
Here we studied the performances of some of the available analytic methods applicable to the analysis of proportion data; namely linear regression, Poisson regression, beta-binomial regression and Generalized Linear Mixed Models (GLMMs). We report the conclusions from a simulation study evaluating power and Type I error rates of these models in scenarios akin to those met by behavioral researchers and differing in sample size, cluster size and fixed effects parameters; plus, we describe results from the application of these methods on data from two real behavioral experiments.
Our results show that, while GLMMs and beta-binomial regression are powerful instruments for the analysis of clustered binary outcomes, linear approximation can still provide reliable hypothesis testing in this context. Poisson regression, on the other hand, can suffer heavily from model misspecification when used to model proportion data.
We conclude providing some guidelines for the choice of appropriate analytical instruments, sample and cluster size depending on the conditions of the experiment.

**Keywords**

*Proportions, Generalized Linear Models, Generalized Linear Mixed Models, Clustered data, Beta-binomial regression, Behavioral science*

**Introduction**

Most physiological parameters studied by biomedical researchers are continuous variables whose distribution approximates well normality; some examples of this are weight, length, height, blood pressure, hormone or protein levels. For this reason, parametric methods assuming normal distribution of the response variable are the most widely used statistical instruments in biomedical research. Data showing strong departure from normality, on the other side, are usually dealt with by transforming them to achieve better Gaussian approximation, or resorting to the use of nonparametric methods. Nonparametric tests, though, have some remarkable weaknesses, such as decreased power and difficulty in dealing with interaction effects; these limitations suggest the use of more powerful methods when they are available. Furthermore, in some fields of research it is not uncommon to see variables arise whose behavior, while not approximating normality, is well described by other known probability distributions.

In behavioral sciences this kind of non-Gaussian behaviors arise naturally quite often, due to the peculiar nature of the measured responses. One example of this are the outcomes of animal or human experiments involving decision making tasks, in which the subject has to choose among two or more different behaviors, with one response being considered a "success" and the other(s) a "failure". In cases like this, the outcome of interest is the ratio of correct choices on the number of trials; we are thus talking of proportion data.

The distributions of proportions usually do not approximate normality; they are generally asymmetrical and only admit a range of values from 0 to 1, with the range of measurements allowed for a Gaussian distribution going from -∞ to +∞ instead. One way to fix this issue is to apply the arcsine square root transformation to the data, but this approach has been shown to provide only minor or no improvements in power over the analysis of untransformed data, while at the same time giving rise to issues of interpretability of results, thus its use is not advised (Jaeger, 2008; Warton & Hui, 2011). Since the outcomes of behavioral experiments such as the decision making tasks we mentioned are a series of Bernoulli trials, logistic regression can be proposed as a more formal solution to their analysis. Despite this, the very common problem of overdispersion, i.e. the excess variance not accounted for by the model (usually due to overlooked sources of variation, such as inter-individual variability, litter and cage effect, etc.), can make the choice of the appropriate analytical instrument and experimental design very challenging to the researcher.

The aim of this paper is to review some of the most relevant methods available for the analysis of proportions in the usual behavioral experimental set-up, and compare their power and type I error rates through simulation, in order to provide guidelines to the reader for the choice of adequate analytic instruments and sample sizes under different experimental designs. We will subsequently show the results obtained from the application of these methods on two datasets coming from actual behavioral experiments.

**Dealing with proportions**

*Linear approximation and binomial model*

Let's consider a behavioral experiment in which $N$ subjects are exposed to a predetermined number $n$ of trials, and in each trial they are required to choose between two different possible responses. Some examples of this kind of experiments are questionnaires; escape tests, in which the subject has to choose the appropriate response to avoid an aversive stimulus; or risky decision making tasks based on reward, used to assess the preference of the subject for "safe" versus "risky" rewards. We may want to assess whether an experimental variable, such as a genetic trait or a drug treatment, has a significant effect on the propensity of the subject towards one of the two choices.

As we noted before, the most widespread method used to deal with such results in behavioral science is linear approximation. This approach is not plain wrong and in fact it can produce acceptable inferences, but its efficiency depends strictly on the goodness of the Gaussian approximation for the data.

A more formal approach, that takes into greater account the nature of the data generating process, consists in considering each of the n × N trials as a Bernoulli process with two possible outcomes, "success" and "failure", with probability of success π and probability of failure 1 - π ; in this case the number of correct responses $y$ is a random variable with a binomial probability distribution of parameter π (Jaeger, 2008).

This approach to the data requires us to perform the analysis using Generalized Linear Models (GLMs), that allow us to model relations between the covariates and the response variable when the latter's distribution is described by a noted non-Gaussian probability function.

*Poisson regression*

One possible alternative to linear regression that takes more into account the data generating process is Poisson regression. Indeed, the so called "law of rare events" states that, when $n$ is large compared to π, i.e. successes are "rare", the binomial distribution approximates the Poisson distribution.

Poisson regression is the optimal solution to deal with count data that can be interpreted as the outcome of a binomial process with an infinite number of trials and a finite number of successes; e.g., when considering the number of occurrences of a certain event in a given amount of time (Cameron & Trivedi, 2013).

In our hypothetical experiment, the single subject is the statistical unit, and the raw number of correct choices it makes is the response to be analyzed through Poisson regression, i.e. we deal with count data instead of proportions. In this case, the model will be expressed in the form:

$$\log(\mathrm{E}(y|x)) = \beta_0 + \boldsymbol{\beta}'\mathbf{x} \qquad (1)$$

Where $\beta_0$ is the intercept and $\boldsymbol{\beta}'$ the vector of fixed effect coefficients. This method is very easy to apply in most statistical software and in particular R; plus, it can be used also to model situations in which $n$ varies from cluster to cluster in the $N$ clusters by including an $offset = \log(n_i)$ into the model. Nevertheless, the efficiency of Poisson regression in a context in which $n$ is limited is hindered by the upper bound on the number of possible correct responses, since Poisson distribution allows for all integer values in the range going from 0 to +∞. Therefore, the Poisson model is misspecified for proportion data. We can expect Poisson approximation to work well only when we have large $n$ and comparatively low π; in fact it has been shown that it can be a powerful alternative to linear regression even in experimental conditions where an upper bound is present (Lazic, 2015), and it has been applied to the study of complex decision making (Giang & Donmez; Paserman, 2016), gambling (James, O'Malley, & Tunney, 2016) and perseverative behavior (Lazic, 2015). Another issue the experimenter might meet when applying Poisson regression is the inflation of Type I error rate in presence of overdispersion; indeed, inference in Poisson regression is heavily dependent on the assumption of equality of mean and variance. However, this problem can be fixed by applying robust sandwich standard error estimators to the model (the so called Huber-White estimator) (Silva & Tenreyro, 2006; White, 1980).

*Beta-binomial regression*

Beta-binomial regression has been proposed as an alternative to linear regression to model clustered binary data, such as in the case of the proportion of successes and failures over a definite number of trials. This model assumes that the response variable follows a beta-binomial distribution, in which overdispersed binomial data are handled by letting the π parameter of the binomial vary randomly, following a beta distribution. The resulting distribution has probability mass function:

$$\binom{n}{y} \frac{\mathrm{B}(y+\alpha, n-y+\beta)}{\mathrm{B}(\alpha, \beta)} \qquad (2)$$

Where α and β are shape parameters and B is the beta function:

$$\mathrm{B}(x,y) = \int_0^1 t^{x-1}(1-t)^{y-1} dt \qquad (3)$$

The use of beta-binomial regression in behavioral and biomedical research in general has been for some time outside of most experimenter's reach because of lack of appropriate software implementation, but is now easily accessible using various statistical software, e.g. the R packages *aod* (Matthieu Lesnoff & Lancelot, 2012) or *aods3* (M Lesnoff & Lancelot, 2013), and its suitability to the analysis of proportions has been assessed previously in various contexts (Crowder, 1978; Hilbe, 2013; Muniz-Terrera, Hout, Rigby, & Stasinopoulos, 2012).

*Generalized Linear Mixed Models*

The fifth available method we take into account are Generalized Linear Mixed Models (GLMM$_S$); in particular, we will consider mixed effects logistic regression with random intercepts.

The most correct way to deal with binary outcomes when the assumption of independence of the observations is met is the ordinary logistic model. This consists in modeling the response variable as the logarithm of the ratio between the probability of success $\pi$ and the probability of failure $1 - \pi$; the *log-odds* or *logit*, as follows:

$$\text{logit}(E(y|x)) = \log \frac{\pi}{1-\pi} = \beta_0 + \boldsymbol{\beta}'\mathbf{x} \quad (4)$$

However, this cannot be applied straight away in presence of clusters of more or less highly correlated observations, such as in the case of repeated measures on the same subject. One way to fix this issue is using GLMMs. This allows us to model the inter-cluster variability by fitting a mixed effect model with "cluster" as random effect and a series of $n_i$ binary outcomes for each $i^{th}$ cluster in $i = [1,...,N]$. In our hypothetical experiment, we will have $N$ subjects with $n_i$ trials for each $i^{th}$ subjects. Since we expect the observations made on the same subject to be correlated, we account for the overdispersion due to inter-individual variability by allowing the model to fit a different intercept for each $i^{th}$ subject.

In this case, the model will have the form:

$$\text{logit}(E(y|x)) = \log \frac{\pi}{1-\pi} = \beta_0 + \boldsymbol{\beta}'\mathbf{x} + \mathbf{b}'\mathbf{z} \quad (5)$$

Where **b'** is the vector of coefficients for the random effects and **z** is the vector of the cluster-specific random effects. This is perhaps the most formally rigorous analytic method to model the kind of data we described (Jaeger, 2008), but it is also somewhat complex to approach. In particular, if:

$$b \sim_{i.i.d.} N(0, \sigma_b^2)$$

Then the likelihood for a model of this form is:

$$L = \int \prod_{i,j} f_{Y_{i,j}|Z_i}(y_{i,j}|x_{i,j}, b_i) f_{b_i}(b_i) db =$$

$$\prod_i \int_{-\infty}^{+\infty} \frac{e^{(\beta_0 + \beta' x_{i,j} + b_i z_i)}}{\sum_j e^{(\beta_0 + \beta' x_{i,j} + b_i z_i)}} \frac{e^{2\sigma_b^2}}{\sqrt{2\pi\sigma_b^2}} db_i \quad (6)$$

And its maximization requires dealing with intractable multidimensional integration, which can only be approached through numerical approximation. There are various methods and software packages for GLMMs fit able to deal with this issue (For review see Bolker et al., 2009). Here we will concentrate only on two of the most widely used estimation techniques: Penalized Quasi-Likelihood (PQL) and Laplace method.

PQL is probably the most commonly utilized instrument for GLMM fitting. The theoretical approach behind PQL consists in replacing the Likelihood function with a Quasi-Likelihood function which shares some properties with the true Likelihood, which is then integrated using Laplace approximation and maximized (N. E. Breslow & Clayton, 1993).

Another alternative is to approximate directly the solution to the integrated Likelihood Function with Laplace method (Raudenbush, Yang, & Yosef, 2000). This is more computationally intensive than PQL and is more prone to give issues of numerical stability, but it can theoretically give more accurate results (N. Breslow, 2004). Both methods are implemented in advanced statistical software and in R.

In the following sections, beside using simulated data to assess the efficiency of the different analytical instruments we have listed, we will also compare the performances of these two distinct methods for GLMM fit.

**Simulation study**

In order to asses power and Type I Error rate of the different methods we listed in inference on fixed effect, we simulated data from a random intercept logistic model (5) under the following conditions:

- Three sample sizes: 16, 24, 32;
- Random intercepts generated from a centered Gaussian distribution with $\sigma$ = 0.5 or 1.75, representing "weak" and "strong" overdispersion;
- Four cluster sizes: 30, 15, 8, and varying from cluster to cluster with min = 1 and max = 30;
- Two fixed factors, "factor 1" and "factor 2", with the coefficient for factor 1, $\beta_1$, ranging from 0 to 3 (or saturation of statistical power), and the coefficient for factor 2, $\beta_2$, fixed to 0.

Cluster number and size have been chosen as they are reasonable conditions in usual behavioral experiment designs. The two $\sigma$ for the distribution of random intercepts were chosen since they give rise, respectively, to a bell-shaped approximate normal distribution and to an almost-uniform distribution over the (0,1) interval for the probability of correct response $\pi$. Datasets with varying cluster size were generated by assigning to each cluster of size 30 a different probability of non-response, the overall mean probability of non-response being 0.5. Factor 2 was introduced in the models in order to simulate the effect of an ineffective treatment on hypothesis testing.

We simulated 1 000 datasets for each combination of parameters. The following models were fitted to the data: linear regression to counts (or proportions, with *n* non constant); mixed effect logistic regression to clustered binary outcomes (fitted through PQL or Laplace method); Poisson regression to counts; beta-binomial regression to the *correct/incorrect* response ratios. In Poisson regression with non-constant cluster size an offset equal to log(*n*) was included into the model.

We tested the hypothesis of the fixed effect coefficients $\beta_1$ and $\beta_2$ being different from 0 by using Wald's test to compute p-values, and fixing the significance level at α = 0.05; in mixed logistic regression degrees of freedom were calculated using Satterthwaite approximation; in Poisson regression, both ordinary and robust standard error estimators were used for calculation of p-values. P-values significant at α = 0.05 were counted to generate power curves; we also calculated Type I Error rates by counting significant p-values when both coefficients were = 0. When the models showed high rates of non-convergence or computational errors (>10%) the scenario was discarded. The results of the analysis are reported in **Fig. 1, Fig. 2** and **Table 1**.

The entire analysis was performed on R (RCoreDevelpmentTeam, 2014), in particular using packages *MASS* (Venables & Ripley, 2002), *lme4* (Bates, Mächler, Bolker, & Walker, 2014) and *aod* (Matthieu Lesnoff & Lancelot, 2012).

**Example 1: Escape Test**

The first example we will deal with is the avoidance test for Escape Deficit (ED), an instrument for the evaluation of Learned Helplessness Syndrome used to assess the ability of the animal to develop an avoiding response to repeated aversive stimuli.

In this test the animal is placed in an apparatus consisting of a cage with dark walls, divided into two equal chambers by a dark partition with a sliding door; one half of the cage is connected to a generator able to deliver weak electric shocks to the animal (the electric chamber), while the other is not (neutral chamber, **Fig. 3A**).

After an habituation period in the apparatus, the animal is placed in the electrified chamber and receives a series of few-seconds long electric shocks, usually at a 30 sec intervals from each other, in coincidence with the opening of the door connecting the electrified chamber to the neutral one. In normal conditions, the animals easily learn to avoid the aversive stimulus most of the times by moving to the neutral chamber when the shock is delivered. After each trial, animals succeeding in escaping are gently placed again in the electrified chamber and the procedure is repeated a predetermined number of times.

In this kind of experiment, we have a dichotomous outcome (escape/failure) which is repeated $n$ times for each of the $N$ animals, and we want to assess whether animals administered with different treatments show a significantly different performance in this task. We reanalyzed a dataset from a study by Scheggi et al. (Scheggi, Pelliccia, Ferrari, De Montis, & Gambarana, 2015), coming from an ED test in which rats ($N = 17$) were assigned to three experimental groups and tested in the ED paradigm. 6 animals were subjected to a chronic stress protocol prior to the experiment, 6 animals were subjected to stress and treated with a classic antidepressant drug (Imipramine), and 5 animals were treated with saline solution and acted as controls. Each animal received 30 electric shocks, and the responses, escapes or failures, were registered.

We tested the hypothesis that animals in the treated groups have an ability to enact the avoiding response significantly different from the one of control animals. To this aim we used one-way ANOVA, Poisson regression, beta-binomial regression and GLMMs.

The results are reported in **Fig. 3B** and **Table 2**. Poisson regression stood out in this example since it gave very low p-values, but the high residual deviance/residual degrees of freedom ratio (4.536357) suggested an artifact due to overdispersion; in fact the use of robust standard errors gave much more reasonable p-values (**Table 2**).

**Example 2: Preference test**

The data for our second example come from a study from Marchese et al.(Marchese, Scheggi, Secci, De Montis, & Gambarana, 2013) about the effect of stress and long-term treatment with lithium on the acquisition of operant behavior in rats. In a similar fashion as in example 1, here experimental subjects were assigned to three groups (controls, stressed, and stressed treated with lithium, for a total $N = 6 + 6 + 8 = 20$), and tested for their competence to acquire a vanilla sugar (VS)-reinforced instrumental behavior [VS-sustained appetitive behavior (VAB)] in a Y-maze preference test paradigm. In each trial, one animal was placed in the starting arm of an Y-maze (15 × 40 × 20 cm for each arm, **Fig. 3C**) with a reward (a pellet of vanilla sugar) in one of the divergent arms. Each subject underwent ten daily trials, and in each trial one of three possible responses was recorded: *correct response*, if the animal moved into the arm containing the reward; *incorrect response*, if the animal went into the empty arm; *incomplete trial*, if the animal stayed in the starting arm until the end of the trial.

In the original work, the experiment was repeated daily for ten days in order to study the ability of the animals to learn the instrumental behavior, and the variation in time of the number of correct responses, incorrect responses and incomplete trials was analyzed with repeated measures ANOVA. Yet, this kind of design includes two potential levels of clustering of the binary outcomes (subjects and days), which goes beyond the scope of our study; thus, we only analyzed the outcomes from day 2. This was chosen since in the study by Marchese et al. differences between the groups were not yet detectable at this point in time. Here we did not look into the number of incomplete trials, but only into the proportion of *correct/incorrect* responses. Results of the analysis are presented in **Fig. 3D** and **Table 3**. In this example, robust Poisson regression was the only method detecting a difference between the control and stress groups.

**Discussion**

Even though the non-normality of the distribution of proportion data may not be an issue when dealing with a sizeable number of observations, such as in the case of, e.g., phase III clinical trials, it is likely to become a serious confound in conditions where the number of observation is more limited. This is typically the case for behavioral studies in psychology or cognitive science, that often rely on small samples only adequate to the detection of large effect sizes (Marszalek, Barber, Kohlhart, & Holmes, 2011). Small sample sizes become a compelling issue, also, in preclinical drug research on animals: in this case, ethical and economic concerns alike call for the minimization of the number of subjects used in the experiments; furthermore, in experiments where the repeated trials are a source of stress for the subject (e.g., the escape tests), it is in the experimenter's interest to minimize confounding effects due to excessive stress. In these conditions, the choice of appropriate analytic instruments, sample size and number of trials are pivotal for the study to answer its experimental questions; on the other hand, the usual methods for sample size calculation are not entirely adequate to deal with non-normally distributed clustered data.

One advisable alternative is to perform preliminary power analysis through simulation (Johnson, Barry, Ferguson, & Müller, 2015; Kain, Bolker, & McCoy, 2015), but this is time consuming, technically challenging, and it must be tried on a wide predetermined range of scenarios in order to determine a power curve. For this reason, in this work we wish to provide some practical guidelines to the choice of the appropriate statistical instruments and experimental design, in some scenarios that can reasonably be met by behavioral researchers dealing with heavy limitations on sample and cluster size.

Among the statistical methods taken into consideration in our study, linear approximation performed quite well on proportion data, showing a type I error rate consistent with the nominal 5% level of significance, or below. Nevertheless, it was also more conservative compared with GLMMs and beta-binomial regression. As expected, this gap in power became less and less significant as the number of observations increased, and at $N = 24$ and $n = 30$ it was quite marginal. The unremarkable gain in power granted by the use of GLMMs or beta-binomial regression when both $N$ and $n$ are big enough, considering the relatively challenging nature of these methods, may advise against their use when the above conditions are met.

Poisson regression performed quite poorly in our study, both applying ordinary and robust standard errors estimates. This was particularly evident when we applied ordinary Poisson regression to datasets with strong overdispersion ($\sigma = 1.75$), where we found Type I error rates as high as 43%. The issue was easily fixed when sandwich standard error estimators were applied, but this came at the cost of a remarkable loss in power compared to other methods. We therefore conclude that Poisson regression, even with robust standard error estimators, should not be seen as a first choice when dealing with proportion data, and that better alternatives are available even when cluster size is constant but moderate.

Beta-binomial regression was shown to have some interesting properties. While having a higher Type I error rate compared to linear approximation, it was also consistently more powerful; this was very evident in "worst case scenarios", with small or non-constant $n$, small $N$ and strong overdispersion.

For what concerns logistic GLMMs, a noticeably different outcome between fit performed through PQL and Laplace approximation was found. PQL was generally more powerful compared to linear approximation, but not to beta-binomial regression. Nevertheless, the gain in power compared to linear regression was quite thin, being as small as 0.2% in some conditions, namely large sample size, small cluster size and weak overdispersion. This was balanced by a very acceptable Type I error rate, spanning from a minimum of 3.8% to a maximum 8.0%.

When parameter estimation was performed through Laplace approximation, the logistic mixed model was generally the most powerful method (with the exception of simple Poisson regression, which also gave unacceptable Type I error rates), but it also had Type I error rates consistently above the nominal 5%. Also, in some scenarios, namely small or non-constant $n$, small $N$, strong overdispersion and high $\beta_1$, this method was prone to give serious issues of non-convergence. PQL, while being overall less powerful than Laplace, was somewhat less prone to this kind of problems. This is consistent with previous findings from simulation studies of multilevel logistic models comparing different methods for GLMM fit (Kim, Choi, & Emery, 2013; Moineddin, Matheson, & Glazier, 2007).

In this respect, Moineddin et al. (Moineddin et al., 2007) and Kim et al. (Kim et al., 2013) studied the performances of different methods for logistic GLMM fit in parameter and standard error estimation. We, on our part, did not look into properties of parameter estimation *per se*, focusing on hypothesis testing instead. This choice was dictated by the demands of preclinical research with limited sample sizes, that rarely allow for the detection of very small effect sizes, therefore making the issue of detectability a priority.

For what concerns sample and cluster size, we observed that in presence of weak overdispersion increasing sample size had similar effects on power to increasing cluster size. The most relevant improvement in this sense was observed in the shift between $N = 16$ and $N = 24$, and between $n = 8$ and $n = 15$. On the other hand, as expected, in presence of strong overdispersion changes in cluster size had little effect on power, whereas increasing sample size had more relevant consequences; e.g., using beta-binomial regression and a $\beta_1 = 1.5$, there was only an increase of roughly 4 percent points in power by increasing $n$ from 8 to 15, whereas the improvement was of about 8 points when increasing $N$ from 16 to 24.

We also compared the performances of the models on two real datasets from actual behavioral experiments on rodents, namely escape test and Y-maze preference test. In the first case, we had a small sample size of 17 subjects and a large cluster size of 30 trials, in the second case we had a moderate sample size of 20 subjects and a non-constant cluster size with a maximum of 10 completed trials. Results from example 1 are mostly consistent with simulation outcomes; in fact, the mixed logistic model fit through Laplace method and beta-binomial regression gave the lowest p-values. Poisson regression gave even lower p-values, but this result was likely a false positive due to unaccounted overdispersion. Indeed, we ran a test for overdispersion on the Poisson model from example 1 using the *dispersiontest* function from R package *AER* (Kleiber & Zeileis, 2008), discovering that there was indeed strong evidence for overdispersion (p-value = 0.003669).

Some of the results from these analyses seemed, at first sight, inconsistent with the outcomes of simulation; in particular, in the VAB experiment Poisson regression with sandwich estimators was the only method able to detect a difference between groups. This may seem puzzling, since in simulated scenarios with non-constant cluster size robust Poisson regression was not the most liberal analytic method. Yet, it has to be considered that the Poisson model is misspecified for proportion data; thus, the overdispersion in the logit model generating the data does not necessarily show up as such when a Poisson model is applied. In fact, the test for underdispersion was highly significant in the Poisson model from example 2 (p-value = $6.732^{-16}$). This is not unexpected since Poisson regression is not the most appropriate way to model conditions in which $n$ is not very big compared to $\pi$. In particular, the upper bound to the number of responses is likely to put a strong constraint on variance, resulting in underdispersion and conservative inference. This underdispersion is accounted for when robust standard error estimators are applied.

In fact, plotting the ratio between residual deviance and residual degrees of freedom from the Poisson models applied to our simulated datasets, we find that, when the logit model generating the data is only weakly overdispersed and cluster size is small, application of a Poisson model frequently leads to underdispersion, whereas when the logit model shows strong overdispersion so does the Poisson model (**Fig. 4**).

Indeed, in our simulated scenarios with non-constant cluster size, maximum cluster size was 30, whereas in the VAB experiment it was only 10. Also, if we assume a mixed effect logistic model as the data generating process, the estimate for $\beta_1$ obtained using Laplace method to approximate maximum likelihood is -0.7239. In our simulations, under weak overdispersion, small $N$ and $n$, and $\beta_1 < 1$, robust Poisson regression was the most liberal method, slightly exceeding in this respect even beta-binomial regression and GLMMs.

**Conclusions**

Our study has the obvious constraints of a simulation study, plus a few other limitations.

Firstly, we only studied the performance of random intercept models, thus, our results are not directly applicable to scenarios in which single subjects respond differently to a certain treatment or condition. Another limit of our work is that we only took into consideration one source of overdispersion, individual clustering, thus assuming that each individual subject is independent of the others. In behavioral sciences, though, especially when working on animal behavior, other sources of extra variability often appear, e.g. litter and cage effect. We did not look into the effects of multiple nested and/or crossed levels of clustering, nevertheless these are likely to be relevant in the everyday practice of the experimenter. Furthermore, we did not simulate scenarios in which the assumption of normally distributed random effects and errors are violated.

The repertoire of methods for the analysis of proportion data that we took into consideration in the present study must not be considered in any way as complete or definitive. Different methods and software packages for GLMM fit exist that we have not put on trial, such as Monte Carlo Expectation Maximization (Levine & Casella, 2001) or Gauss-Hermite Quadrature (AGQ) (Clarkson & Zhan, 2002). AGQ in particular is relevant since it is purportedly more accurate than PQL and Laplace, but it was neglected since it is more computationally intensive than Laplace approximation while using basically the same principle. Indeed, Laplace method uses a Taylor expansion of the likelihood function around one point, while AGQ uses multiple points; therefore, AGQ is in a sense a slower but more accurate version of Laplace (Clarkson & Zhan, 2002). In particular, Kim et al. already showed that it performs better than Laplace with small sample sizes, but the difference tends to vanish as sample size grows (Kim et al., 2013).

Another limitations of our study is our choice to apply only fully parametric models, thus neglecting semi-parametric methods for modeling clustered data, such as Generalized Estimating Equations (GEE).

Also, most importantly, we chose to keep a frequentist framework based on classical hypothesis testing, and did not take into account Bayesian methods. Yet, these represent powerful alternative instruments for the analysis of clustered binary responses.

Despite these limitations, our study provides some directions to behaviorists and other scientists working with proportion data and relatively small sample sizes:

- If we expect little to moderate inter-individual variability, the experimenter can change both sample size and cluster size to attain greater power. In conditions where the single trials are stressful or tiresome to the subject, or there is some other reason to minimize the number of trials, a reasonable compromise between sample size and cluster size for a two-factors design may be using about 15 trials on 24 subjects. In case of a significant probability of non-responses, this has to be adjusted in consideration of the expected number of *completed* trials.
- If strong inter-individual variability is expected, it is advisable to have a bigger sample size; 24 subjects are a reasonable choice for a two-factors design. The number of trials, instead, becomes less important, although it is advisable to have at least about 15 completed trials per subject.
- GLMM fit through Laplace approximation is the most powerful method in most scenarios, while being also somewhat more prone to Type I errors. If not much can be assumed about effect size and variance beyond the clustering structure, the logistic mixed model can be seen as a good first choice for analyzing proportion data. This is expected to hold true also when AGQ is used for maximum likelihood estimates. Should issues of non-convergence arise with GLMMs, beta-binomial regression is the most powerful alternative.
- Overall, linear regression is an acceptable method to perform hypothesis testing with proportion data, and is usually consistent with the nominal level of significance. It is in general more conservative than mixed binomial or beta-binomial models, thus it may not be the best choice for the detection of small effect sizes. Also, caution should be exercised in the interpretation of the estimated parameters, since using linear models to deal with proportion data can give rise to nonsensical predictions, such as probabilities outside the [0,1] interval.

- Poisson regression should not be applied straight away to proportion data because of the strong inflation of Type I error rate it can produce when strong overdispersion is present. Robust Poisson regression, on the other side, has a reasonable Type I error rate but is also not very powerful compared to beta-binomial regression or GLMM fit through Laplace method. Yet, it may be advantaged in the detection of small effect sizes when weak overdispersion is expected. This approach requires great caution, though, since with a small number of trials the Poisson model is heavily misspecified, and will likely give rise to biased estimates and nonsensical predictions. In any case, testing for overdispersion and/or underdispersion is essential when applying Poisson regression to proportion data.

**Conflict of interest**

The authors declare no conflict of interests

|  |  | Type I Error rate (%) | | | | | | | |
|---|---|---|---|---|---|---|---|---|---|
| *Coefficient* | | $\beta_1$ | | | | $\beta_2$ | | | |
| *Cluster size* | | n = 8 | n = 15 | n = 30 | n.c. | n = 8 | n = 15 | n = 30 | n.c |
| | | *N* = 16 | | | | | | | |
| **Linear** | σ = 0.5 | 4.2 | 5.1 | 5.9 | 6.3 | 4.2 | 4.8 | 7.2 | 5 |
| **PQL** | | 3.8 | 6 | 7.8 | 7.5 | 3.8 | 5.5 | 8.0 | 5.6 |
| **Laplace** | | 6.2 | 7.8 | 10.6 | 10.6 | 6.2 | 8.6 | 9.9 | 9.1 |
| **Poisson** | | 1.3 | 3.9 | 9.2 | 5.3 | 1.1 | 3.9 | 9.4 | 3.7 |
| **Poisson (robust)** | | 8 | 9.2 | 10.6 | 10.5 | 7.3 | 9.2 | 10.4 | 8.9 |
| **Beta** | | 6.6 | 8.1 | 10.4 | 10.5 | 6.4 | 9.3 | 10 | 8.9 |
| **Linear** | σ = 1.75 | 5.5 | 5.1 | 5.1 | 5.8 | 4.9 | 4.8 | 4 | 6.4 |
| **PQL** | | 6.3 | 6.1 | 6.2 | 7.7 | 5.6 | 5.7 | 7.2 | 7.1 |
| **Laplace** | | 13 | 9.7 | 9 | 12.1 | 12.6 | 10 | 9.5 | 11.6 |
| **Poisson** | | 13.8 | 24.3 | 41.1 | 27.4 | 13 | 28.4 | 41.9 | 25.8 |
| **Poisson (robust)** | | 7.6 | 7.7 | 7.6 | 8.6 | 7 | 7.3 | 7.8 | 8.5 |
| **Beta** | | 8.7 | 8 | 8.5 | 10.2 | 8.5 | 8 | 8.8 | 10.2 |
| | | *N* = 24 | | | | | | | |
| **Linear** | σ = 0.5 | 4.6 | 4.3 | 6.1 | 3.8 | 5.6 | 6.2 | 4.6 | 5.9 |
| **PQL** | | 4.3 | 6 | 7.6 | 5.1 | 5.4 | 5.5 | 6.2 | 7.4 |
| **Laplace** | | 6.5 | 6.2 | 8.5 | 7.9 | 6.7 | 8.9 | 8 | 8.7 |
| **Poisson** | | 1.5 | 3.2 | 9.7 | 3.2 | 1.7 | 3.9 | 7.5 | 5.3 |
| **Poisson (robust)** | | 7.2 | 6.5 | 8.3 | 6.2 | 7.1 | 9 | 7.7 | 8.5 |
| **Beta** | | 6.6 | 6.2 | 8.5 | 7.4 | 6.7 | 8.8 | 8 | 8.7 |
| **Linear** | σ = 1.75 | 3.9 | 5.1 | 5.6 | 4 | 4.6 | 4.2 | 6.8 | 4.7 |
| **PQL** | | 4.7 | 6.2 | 6.5 | 4.5 | 5.1 | 5.3 | 7.9 | 5.8 |
| **Laplace** | | 9 | 8 | 7.9 | 7.2 | 9.8 | 7.8 | 10.6 | 8.1 |
| **Poisson** | | 11.5 | 24.3 | 41.5 | 24.7 | 12.1 | 25.5 | 43 | 26.3 |
| **Poisson (robust)** | | 4.9 | 6.7 | 7.3 | 5.4 | 5.9 | 5.5 | 9 | 7.1 |
| **Beta** | | 5.8 | 7.3 | 7 | 5.8 | 6.5 | 6.1 | 9.2 | 7.2 |
| | | *N* = 32 | | | | | | | |
| **Linear** | σ = 0.5 | 5.1 | 4.9 | 4.7 | 3.7 | 4.1 | 5.1 | 5.1 | 3.8 |
| **PQL** | | 5.2 | 5.6 | 6.5 | 5.1 | 4.2 | 5.8 | 7.8 | 5 |
| **Laplace** | | 7.1 | 6.6 | 5.3 | 6.2 | 5.5 | 6.9 | 6.4 | 7 |
| **Poisson** | | 1.8 | 3.5 | 8.4 | 2.8 | 1.8 | 4.2 | 8.5 | 3.8 |
| **Poisson (robust)** | | 7 | 6.5 | 6.3 | 6 | 5.9 | 6.4 | 7.9 | 6.2 |
| **Beta** | | 6.8 | 6.7 | 6.4 | 5.9 | 5.7 | 6.7 | 7.9 | 6.3 |
| **Linear** | σ = 1.75 | 6.3 | 6.9 | 4 | 3.5 | 6.4 | 5.3 | 3.9 | 6 |
| **PQL** | | 6.6 | 7.9 | 5.4 | 4.5 | 6.6 | 5.9 | 4.6 | 6.7 |
| **Laplace** | | 10.2 | 9 | 6.4 | 6.2 | 10.2 | 7.2 | 5.7 | 8.3 |
| **Poisson** | | 13.4 | 29.3 | 41.8 | 24.9 | 16.2 | 24.4 | 40.2 | 25.8 |
| **Poisson (robust)** | | 7.2 | 7.6 | 5.6 | 4.2 | 7.3 | 6.2 | 4.8 | 6.9 |
| **Beta** | | 7.9 | 8.2 | 6.3 | 5.7 | 8 | 6.7 | 5.2 | 7.7 |

Table 1: Type I error rates in the simulations; 'n.c.' = non-constant.

|  | *Stress* |  | *Stress + Imipramine* |  |
|---|---|---|---|---|
| *p-values* |  |  |  |  |
| **Linear** | 0.0057 | ** | 0.1285 |  |
| **PQL** | 0.0058 | ** | 0.2237 |  |
| **Laplace** | 0.0013 | ** | 0.2005 |  |
| **Poisson** | $8.75 \cdot 10^{-9}$ | *** | 0.0086 | ** |
| **Poisson (robust)** | 0.0037 | ** | 0.0902 | • |
| **Beta** | 0.0009 | *** | 0.0825 | • |

Table 2: p-values from the ED experiment calculated from the models fitted. Significance levels compared with control group: 0.1, • (suspect); 0.05, * (significant); 0.01, ** (highly significant); 0.001, *** (decisive)

|  | *p-values* | |
|---|---|---|
|  | **Stress** | **Stress + Lithium** |
| *p-values* | | |
| **Linear** | 0.194 | 0.380 |
| **PQL** | 0.200 | 0.825 |
| **Laplace** | 0.176 | 0.819 |
| **Poisson** | 0.447 | 0.911 |
| **Poisson (robust)** | 0.039 * | 0.759 |
| **Beta** | 0.204 | 0.745 |

Table 3: p-values from the VAB experiment calculated from the models fitted. Significance levels compared with control group: 0.1, • (suspect); 0.05, * (significant); 0.01, ** (highly significant); 0.001, *** (decisive)

12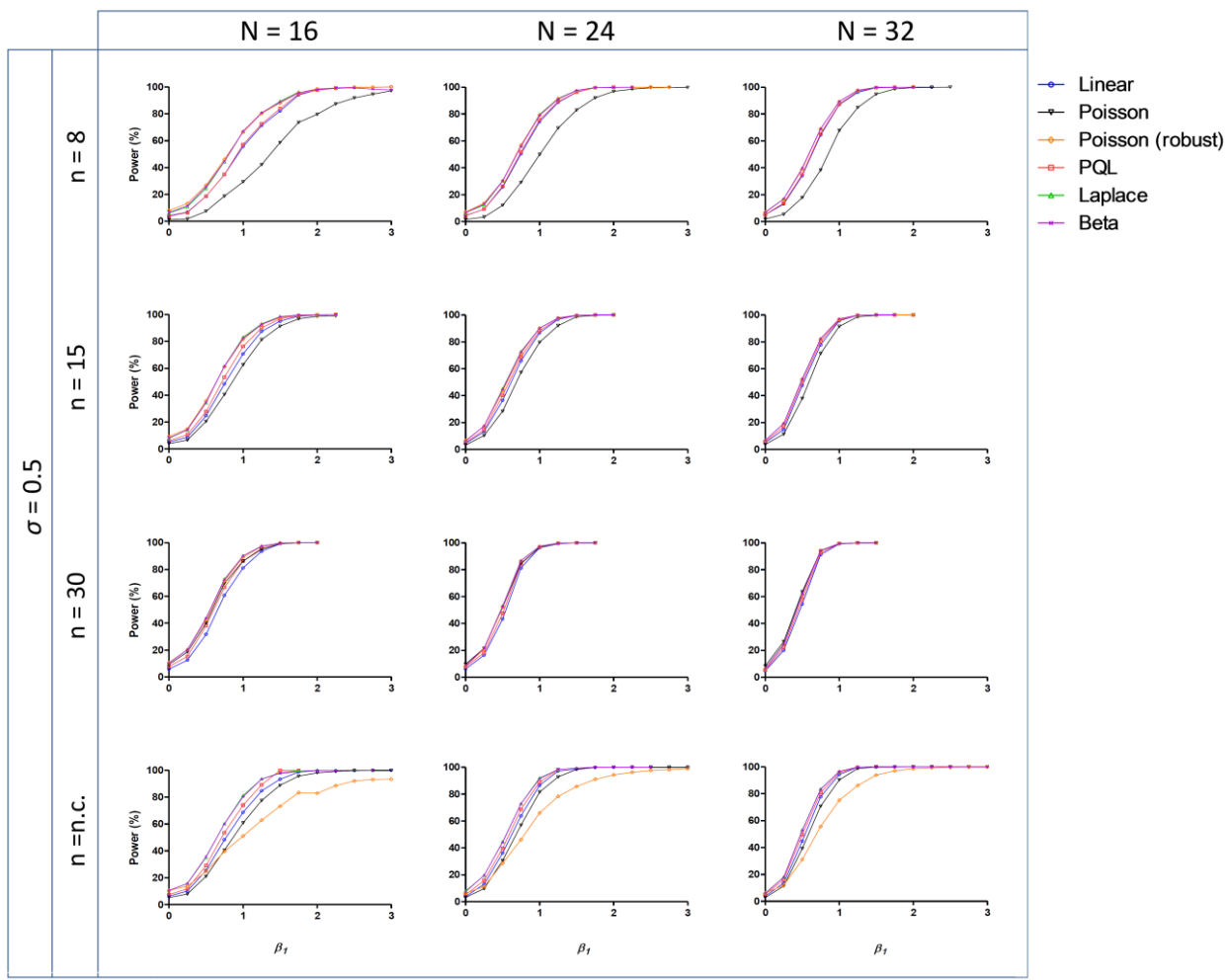

Figure 1: Power curves over 1000 simulations for the twelve different combinations of sample size and cluster size with "weak" overdispersion.

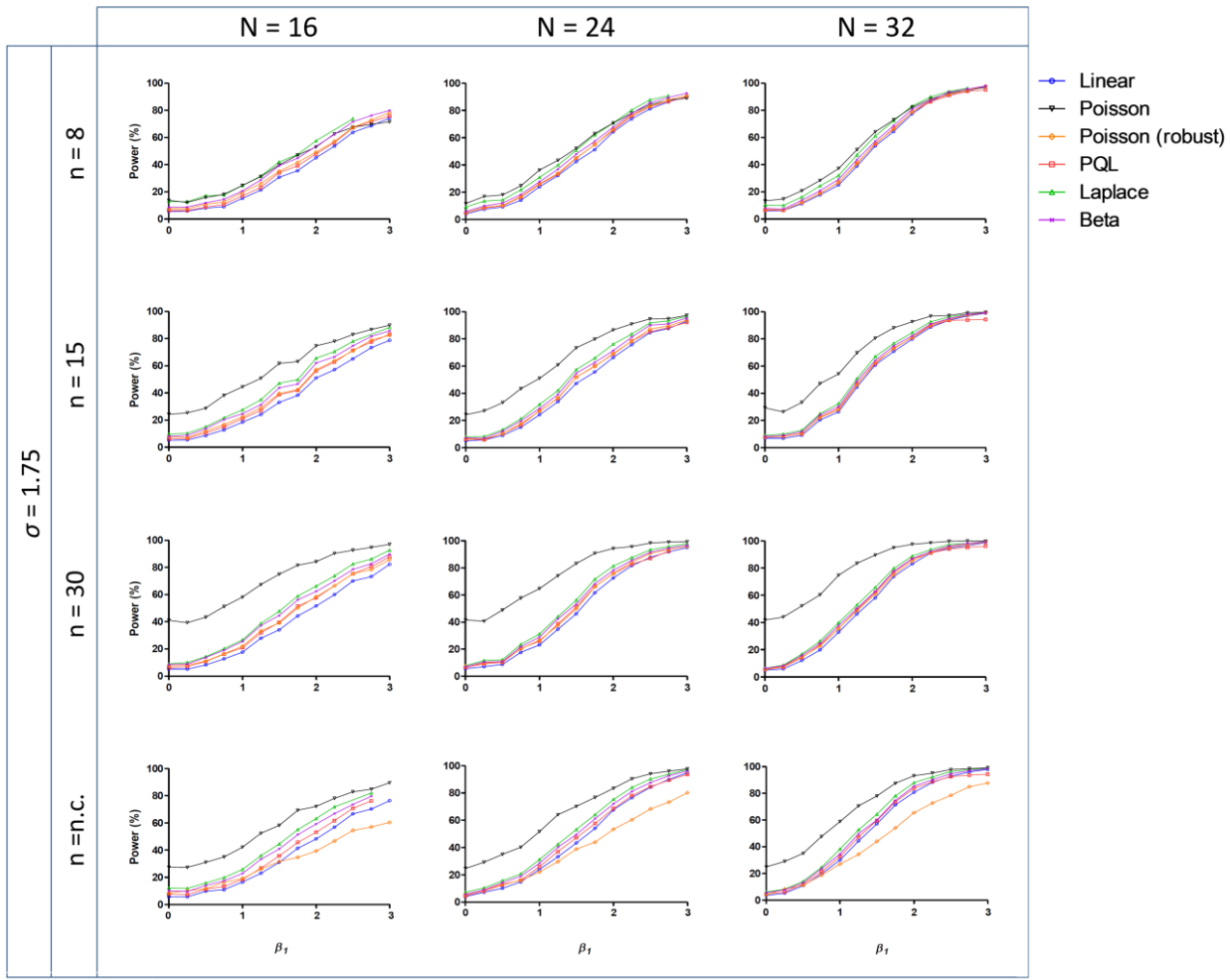

Figure 2: Power curves over 1000 simulations for the twelve different combinations of sample size and cluster size with "strong" overdispersion. Missing points are due to exceedingly high percentages of computational errors in model fit.

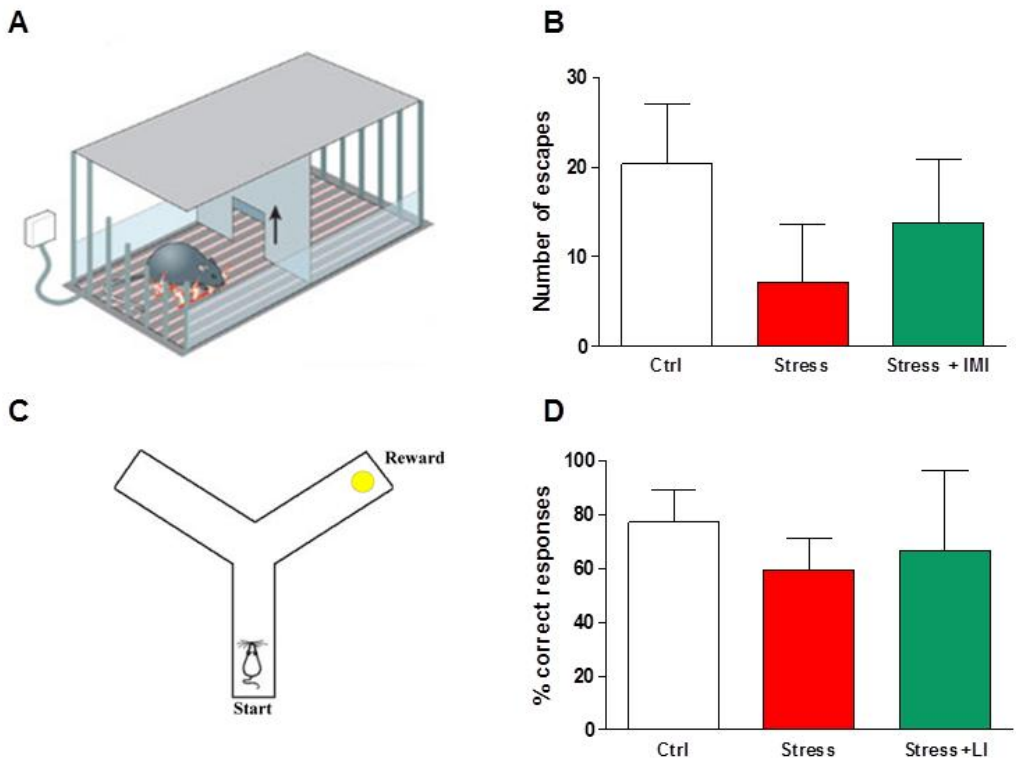

Figure 3: A: Image illustrating the setup of the avoidance learning test (adapted from Krishnan and Nestler, 2008 with permission); B: Number of escapes per experimental group in the ED experiment from Scheggi et al., 2015. Results presented as mean ± SD; C: Image representing the setup for the Y-maze preference test; D: Percentages of correct responses in the VAB experiment from Marchese et al., 2008. Results presented as mean ± SD.

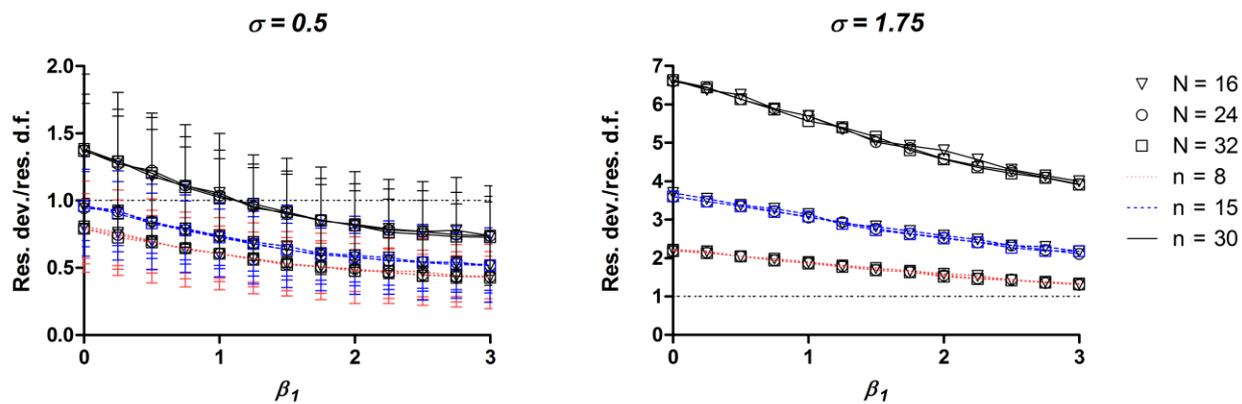

Figure 4: Residual deviance to residual degrees of freedom ratio in Poisson models as a function of fixed effect coefficient, sample and cluster size, and degree of overdispersion of the logit model; the horizontal interrupted line represents ideal equidispersion. Data are presented as mean ± SD over 1000 simulations.